\documentclass[conference]{IEEEtran}
\IEEEoverridecommandlockouts
\usepackage{cite}
\usepackage{amsmath,amssymb,amsfonts}
\usepackage{algorithmic}
\usepackage{graphicx}
\usepackage{textcomp}
\usepackage{xcolor}
\PassOptionsToPackage{hyphens}{url}
\usepackage{url}
\def\BibTeX{{\rm B\kern-.05em{\sc i\kern-.025em b}\kern-.08em
    T\kern-.1667em\lower.7ex\hbox{E}\kern-.125emX}}
\begin{document}

\title{DevPhish: Exploring Social Engineering in Software Supply Chain Attacks on Developers}

\author{\IEEEauthorblockN{1\textsuperscript{st} Hossein Siadati}
\IEEEauthorblockA{\textit{Computer Science Department} \\
\textit{UNCW}\\
Wilmington, USA \\
siadatis@uncw.edu}
\and
\IEEEauthorblockN{2\textsuperscript{nd} Sima Jafarikhah}
\IEEEauthorblockA{\textit{Computer Science Department} \\
\textit{UNCW}\\
Wilmington, USA \\
jafarikhaht@uncw.edu}
\and
\IEEEauthorblockN{3\textsuperscript{rd} Elif Sahin}
\IEEEauthorblockA{\textit{Computer Science Department} \\
\textit{UNCW}\\
Wilmington, USA \\
es5510@uncw.edu}
\and
\IEEEauthorblockN{4\textsuperscript{th} Terrence Hernandez}
\IEEEauthorblockA{\textit{Computer Science Department} \\
\textit{UNCW}\\
Wilmington, USA \\
tbh1597@uncw.edu}
\and
\IEEEauthorblockN{5\textsuperscript{th}  Elijah Tripp}
\IEEEauthorblockA{\textit{Computer Science Department} \\
\textit{UNCW}\\
Wilmington, USA \\
elt8862@uncw.edu}
\and
\IEEEauthorblockN{6\textsuperscript{th} Denis Khryashchev}
\IEEEauthorblockA{\textit{Computer Science Department} \\
\textit{NYU}\\
New York, USA \\
denis.khryashchev@nyu.edu}
\and
\IEEEauthorblockN{7\textsuperscript{th} Amin Kharaz}
\IEEEauthorblockA{\textit{School of Computing and Information Sciences} \\
\textit{Florida International University}\\
Florida, USA \\
amin@iseclab.org}
}
\IEEEoverridecommandlockouts
\IEEEpubid{    \makebox[\columnwidth]{
        \parbox{\columnwidth}{
            979-8-3315-4090-6/24/\$31.00 \copyright 2024 IEEE. Personal use of this material is permitted. Permission from IEEE must be obtained for all other uses, in any current or future media, including reprinting/republishing this material for advertising or promotional purposes, creating new collective works, for resale or redistribution to servers or lists, or reuse of any copyrighted component of this work in other works.
        }
        \hfill
    }
    \hspace{\columnsep}\makebox[\columnwidth]{}}
\maketitle

\begin{abstract}
The \emph{Software Supply Chain} (SSC) has captured considerable attention from attackers seeking to infiltrate systems and undermine organizations. There is evidence indicating that adversaries utilize \emph{Social Engineering} (SocE) techniques specifically aimed at software developers. %The attackers have customized their communication and persuasion strategies to deceive their tech-savvy audience. 
That is, they interact with developers at \emph{critical steps} in the Software Development Life Cycle (SDLC), such as accessing Github repositories, incorporating code dependencies, and obtaining approval for Pull Requests (PR) to introduce malicious code. This paper aims to comprehensively explore the existing and emerging SocE tactics employed by adversaries to trick \emph{Software Engineers} (SWEs) into delivering malicious software. By analyzing a diverse range of resources, which encompass established academic literature and real-world incidents, the paper systematically presents an overview of these manipulative strategies within the realm of the SSC. Such insights prove highly beneficial for threat modeling and security gap analysis.
\end{abstract}

\maketitle



% \fi
%%\IEEEpeerreviewmaketitle

\section{Introduction}
Software development companies and their clients are increasingly alarmed by the prevalence of software supply chain breaches. Prominent instances, including the breach of SolarWinds' build system~\cite{solarwinds-hack}, JumpCloud's and S3CX's compromise of developer environment~\cite{jumpcloud-2023, s3cx-2023}, the exploitation of a critical vulnerability within the widely used Log4j open source dependency~\cite{hiesgen2022race}, unauthorized access to sensitive data via Codecov's container image creation process~\cite{codecov-hack}, and customer breaches stemming from vulnerabilities in Kaseya's VSA software~\cite{kaseya-hack}, have starkly highlighted the potential for widespread disruption resulting from a single supply chain attack.
\newpage
These attacks encompass various intricate steps and often leverage technical expertise, frequently exploiting human vulnerabilities to initiate or progress the attack. In the \emph{event-stream} hack, for instance, attackers lured maintainers of a widely used Open-Source NPM library named \emph{event-stream}. They then assumed the role of maintainers and introduced a new dependency that offered a valuable functionality alongside a hidden backdoor enabling the attackers to steal Bitcoin from e-wallets~\cite{eventstream-hack}. 

Numerous studies have delved deep into its technical dimensions in the expansive domain of SSC security. Yet, a notable gap remains, where the realm of social engineering intrinsic to these incidents remains relatively unexplored. In particular, there exists a lack of research investigating the nuanced interplay of social engineering in leveraging the human element, particularly within the context of SWEs. We have coined the term \emph{DevPhish} to encapsulate the act of manipulating SWEs, inducing them to unknowingly introduce malicious code into the software. This novel term draws attention to the significant vulnerability introduced by human psychology, emphasizing the need to dissect the intricate connections between technological susceptibilities and human manipulation within the SSC landscape.

To address the void in the domain, this research undertakes an in-depth examination of the realm of social engineering within the SSC by analyzing real-world incidents. This study embarks on the exploration of the following paramount questions:

\begin{itemize}

    \item  To what degree do social engineering tactics play a role in the success of SSC attacks?\label {Q1}

    \item  How do attackers interact with SDLC steps to launch social engineering attacks?\label{Q2}

    %\item  What psychological factors and social engineering tactics increase software engineers' vulnerability to social engineering techniques in SSC attacks?\label{Q3}

\end{itemize}

%This study serves multiple objectives. Primarily, it constructs a DevPhishing taxonomy by thoroughly investigating social engineering occurrences within the realm of SSC, intertwining them with SDLC phases. Secondly, it evaluates the impact of social engineering on SSC attacks, drawing from real-world instances. Furthermore, this study sheds light on prevalent social engineering techniques and their predominant utilization points within the SSC. Lastly, it unveils prospective social engineering attacks that have yet to surface, enabling security researchers to anticipate vulnerabilities before they are exploited proactively. 
%by measuring the percentage offers guidance for robust social engineering threat modeling in SSC, arming defenders against these calculated assaults. Ultimately, this research enhances SSC security discussions and empowers defenses against these intricately orchestrated attacks.

The structure of this paper is as follows: Chapter~\ref{ch-ssc-security} provides the required background on SSC Security. Chapter~\ref{ch-research-method} explains our data collection and labeling methodology, and Chapter~\ref{ch-soce-techniques} categorizes attack techniques that involve social engineering. Chapter~\ref{ch-analysis} analyzes attempts at answering our research questions. 
%%%%%%%%%%%%%%%%%%%%%%%%%%%%%%%%%%%%%%%%%%%
\section{Software Supply Chain Security}
\label{ch-ssc-security}
\subsection{Software Supply Chain}
SSC parallels the \emph{stages} of a traditional supply chain:

\begin{itemize}
    \item \textbf{Source Code Development:} Analogous to raw material acquisition, developers create the foundational \textit{source code} that serves as the building blocks of the software.
    
    \item \textbf{Build -- Compilation and Packaging:} Similar to manufacturing and assembly, the \textit{source code} is compiled, processed, and packaged to create the final \textit{software product}. This step involves compilation, integration, and quality checks.
    
    \item \textbf{Distribution:} Just as products are distributed, \textit{software} is distributed through repositories, websites, or app stores. Ensuring accessibility and proper version management is crucial.
    
    \item \textbf{Deployment:} Like retail and point of sale, users \textit{deploying} the software on their devices and infrastructure. This phase initiates user interaction with the software.
    
    \item \textbf{Runtime Support and Updates:} Following deployment, the software system operates on machines, engaging users in interactions with the software. This phase aligns with customers utilizing the service, ensuring system availability, and maintaining the system's currency with the latest feasible versions.
\end{itemize}

\begin{table*}[htbp]
  \centering
  \caption{Summary of DevPhish types}
  \begin{tabular}{|p{4.5cm} p{8cm}|}
    %\toprule
    \hline
    \textbf{DevPhish Type} & \textbf{Description} \\
    %\midrule
    \hline
    Account Compromise & Stealing credentials for developer's platforms, equivalent of traditional phishing (e.g., Phishing github account) \\
    Device Compromise & Compromising developer's device by luring them to install malware (e.g. Installing  malicious IDE extensions)\\
    Malicious Pull Request & Pull request to merge malicious code (e.g., an attacker posing as an open source contributor adding a malicious dependency to code) \\
    Malicious Dependency Watering hole & Malicious dependencies that are inadvertently added to a code (e.g., importing Typosquatting packages) \\
    Malicious Code Snippet Watering hole & Malicious code that are added to code inadvertently (e.g., copied code from StackOverflow) \\
    Entering the Rank of Maintainers & Social Engineering to become an open source project maintainer\\
    %\bottomrule
    \hline
  \end{tabular}
  \label{tab:critical-activities}
\end{table*}

When a feature needs to be implemented, a developer crafts the necessary code to realize this functionality including custom logic written in a specific programming language and dependencies from open-source projects. The developer then initiates a PR, encompassing the new code, and requests a review from a colleague. Typically, this review process involves feedback and iterative revisions until the code aligns with established standards and can be merged into the company's code repository.

Following the merge, an automated process called Continuous Integration/Continuous Delivery (CI/CD) comes into play. This process involves recompiling or rebuilding the software and deploying the output—an application or container image—into the company's registry. Subsequent scripts retrieve these artifacts from the registry, facilitating deployment within the company's computing clusters.

The SSC also encompasses aspects such as public container registries and housing executable components that interface with a software system. Additionally, there are contributing open source developers and distinct roles, including Site Reliability Engineers (SREs) who wield substantial access privileges and often oversee operational services.

\subsection{Software Supply Chain Attacks}

\begin{figure}[!t]
\centerline{\includegraphics[width=8cm,height=7cm,keepaspectratio]{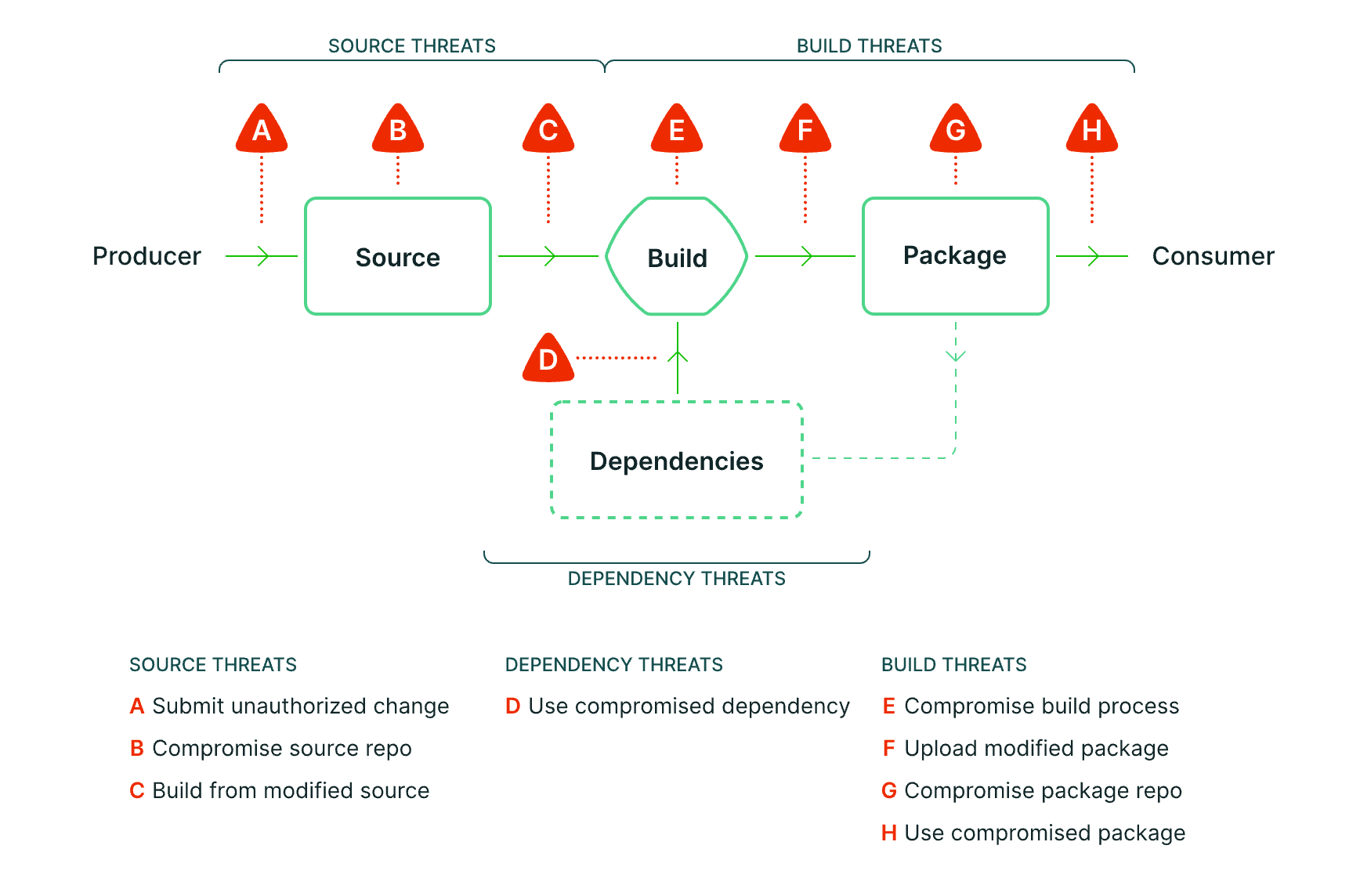}}
\caption{Software Supply Chain steps and threats based on documentations of
SLSA framework~\cite{slsa}}
\label{fig:slsa}
\end{figure}

SSC attacks aim to transform a genuine software artifact into a malicious one by compromising the SDLC. Consequently, software counterfeiting, malware, and Trojans, which are fundamentally unauthorized, do not fall under the classification of SSC attacks.

An instance of SSC attacks involving social engineering is the security breach of JumpCloud, a zero-trust directory platform service utilized for identity and access management. In this incident, a North Korean threat actor successfully executed a spear-phishing attack on a developer, leading them to download malicious code onto their device unwittingly. This provided the threat actor with developer-level access to JumpCloud environments, explicitly targeting their \emph{commands framework}. This framework was subsequently exploited for injecting malicious data into customer environments~\cite{jumpcloud-2023}.

%A prime illustration of such an attack is the SolarWinds breach, which came to light in late 2020. In this instance, sophisticated attackers compromised SolarWinds' software build system (i.e., Compilation and Packaging), embedding a backdoor on the fly to the code that was built. The tainted update was distributed to thousands of organizations using \emph{Orion platform}, including government agencies and corporations. The scale of the breach was massive, with around 18,000 organizations downloading the malicious update, and a few hundred were impacted. This includes at least nine U.S. federal government agencies, including Department of State and Department of Homeland Security (DHS). 

%The particularly concerning aspect of the SolarWinds attack lay in its extended period of undetection, lasting several months—estimated between 9 to 12 months—before discovery. This prolonged duration provided the attackers with substantial time to engage in espionage activities and extract sensitive information, significantly amplifying the attack's impact and severity
%What made the SolarWinds attack particularly alarming was the duration it went undetected for several months before being discovered, spanning approximately 9 to 12 months. This gave the attackers ample time to conduct espionage and exfiltrate sensitive data.

The primary framework for delineating the scope of this attack is the Supply Chain Levels for Software Artifacts (\textbf{SLSA})~\cite{slsa}, initiated by Google and established within the Open Source Security Foundation. While this framework assists in identifying SSC threats (see Figure~\ref{fig:slsa}), it lacks a detailed perspective for modeling social engineering within the SSC.

\subsection{Scope of work}
\begin{figure}[!t]
\centerline{\includegraphics[width=7cm,height=5.5cm,keepaspectratio]{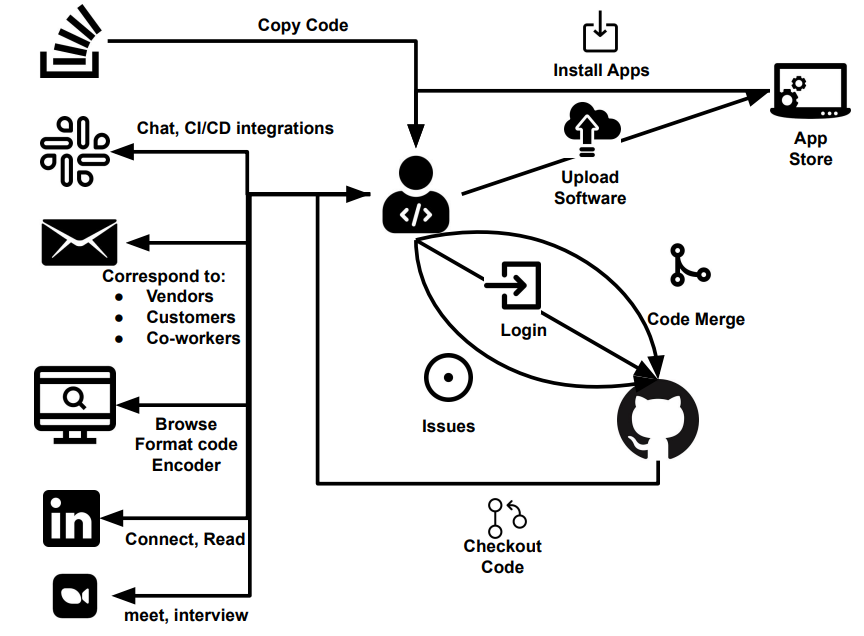}}
\caption{Example interactions between software developers and Software Development Life Cycle (SDLC) critical steps}
\label{fig:swe-critical-activities}
\end{figure}

The primary emphasis of this study is DevPhish for SSC attacks, wherein developers are subjected to social engineering during their direct involvement in the software development life cycle. Figure~\ref{fig:swe-critical-activities} illustrates examples of these interactions.

\section{Research Methodology}
\label{ch-research-method}

The methodology employed for this study involved conducting a Systematic Literature Review (SLR) to carefully select and evaluate pertinent studies from a vast body of existing literature, including academic papers, industry reports, and real-world incidents. In addition, relevant entries from news outlets and bug bounty reports were incorporated to supplement our research. To find these resources, we used keywords such as \emph{software supply chain attacks}, \emph{social engineering software supply chain}, and \emph{software engineer social engineering} on both the Google search engine and Google Scholar. We also used a major listing containing SSC attacks compiled by domain experts~\cite{oss-ssc-attack-incidents-2023}. The result is a list of 198 unique SSC attacks. Afterward, we categorized the attacks into buckets based on their \emph{first point of impact} on developers. We used multiple labeling and cross-validation to make sure the labeling aligns between a team of 4 individuals.

\section{Social Engineering Techniques in SSC Attacks}
\label{ch-soce-techniques}

We have classified the DevPhish techniques employed in SSC attacks, which are based on social engineering, into six categories outlined in Table~\ref{tab:critical-activities}. Each category encompasses various variations.
%Figure~\ref{fig:swe-critical-activities} illustrates the attack activities in different steps of software supply chain lifecycle. 
%We observe that social engineering attacks target the critical activities of software engineers. 
%These activities are illustrated in Figure~\ref{fig:swe-critical-activities} and described in Table~\ref{tab:critical-activities}.
%This section categorizes different types of DevPhish based on the method of attack and stage of SDLC in which the attack happens. This originates from real-world attacks and those identified and deliberated within the security community. Such sources were found based on what is described in Section~\ref{research-method}. 
%\ak{does the figure point to the right figure. seems Figure 1 is more relevant. }
%We use the following abbreviations are use through this section:
%\begin{itemize}
%    \item SCD: Source Code Development 
%    \item VCS: Version Control System 
%    \item MC: Malicious Commit
%    \item MD: Malicious Dependency
%\end{itemize}
%\textbf{add a table for all above}
%\textbf{add a table with a breakdown of the attack and how it happened}

\subsubsection{\textbf{Account Compromise}} Compromised accounts are a significant vector for cyberattacks, with phishing being a common strategy employed when attackers' attempts at guessing attacks fail. While it might seem less probable to deceive someone knowledgeable in technology into surrendering their credentials, human psychology prevails when executing a meticulously crafted attack. As an illustration, Dropbox experienced a data breach resulting from an attack that targeted its employees through phishing emails. 
These phishing emails imitated CircleCI~\cite{dropbox-github-repos-compromise}, a CI/CD platform essential to their Software Development Life Cycle (SDLC), resulting in the compromise of employee accounts. Subsequently, the attackers successfully exfiltrated 130 Github repositories containing vital business information. Notably, this attack was related to a larger GitHub phishing campaign~\cite{github-phishing-campagin-22} that aimed at a broader group of developers.

In addition to stealing business-critical information, compromised developer accounts are frequently exploited to inject malicious code into source codes. A notable incident involves the compromise of the official PHP Git repository, where attackers submitted code changes to the PHP source code. These commits were falsely attributed to recognized PHP developers and maintainers~\cite{php-gitcompromise}. This compromise impacted two million websites utilizing this package. Another example is \emph{ua-parser-js} hack where the attackers compromised a developer account and uploaded malicious versions of software~\cite{ua-parser-compromise}.

Developers manage multiple accounts to access various platforms essential for their daily tasks. These platforms encompass those dedicated to software development (such as Github, cloud platforms, and repositories) and communication tools (such as Zoom and Docs), as illustrated in Figure~\ref{fig:swe-critical-activities}. Of particular interest to attackers are the package repository accounts. In pursuit of this, attackers focus on developers' accounts on package repositories like PyPI and NPM, where they upload malicious versions of legitimate packages. Noteworthy instances include a phishing campaign explicitly designed to steal PyPI developers' credentials~\cite{pypi-phishing}. 

The persuasion techniques used by attackers align with developers' anticipated messaging. For example, a phishing message claims there is a new process for authentication due to "a surge in malicious packages being uploaded to the PyPI.org domain"~\cite{pypi-phishing}.

\subsubsection{\textbf{Device Compromise}} 
Another form of phishing targeting developers involves tricking them into installing malware on their devices. Regardless of how the malware manages to infiltrate a developer's machine, the repercussions are severe. For instance, in the case of the CircleCI breach~\cite{circle-ci-hack-23}, an employee fell victim to an elaborate spear-phishing scheme, resulting in the download of an application supposedly for viewing/signing a critically important PTX document. The compromised device then granted unauthorized access to customers' secret tokens, which were subsequently exploited for information theft. In the JumpCloud breach, a highly skilled North Korean threat actor successfully executed a spear-phishing attack on a developer, leading them to unknowingly download malicious code provided by JumpCloud onto their device. This granted the threat actor developer-level access to JumpCloud developer environments~\cite{jumpcloud-2023}.

In a different iteration of this attack, the compromise of a device occurs during the installation of package dependencies, wherein a malicious pre/post-installation script infiltrates the device. In essence, an attacker could encapsulate a legitimate package with a deceptive installer. As a result, the code itself remains uncompromised, but the developer's machine falls victim to the compromise. In a specific instance, attackers exploited the NPM package installation process~\cite{npm-developer-trojan, npm-malicious-sehll} to deceive developers who only inspect the package's content.

%In the NotPetia incident~\cite{NotPetya-phishing}, attackers used phishing emails to take over MeDoc accounts, eventually installing a backdoor on their update server from which they deployed MeDoc software containing NotPetya — everywhere. 

%Account compromise also stands out as the primary avenue to initiate lateral movement and compromise of the build systems. A notable case is the recent North Korean attack involving JumpCloud, believed to have commenced with a spear-phishing campaign aimed at pilfering credentials. Another instance pertains to the CCleaner attack \cite{ccleaner-attack}, where the breach originated from an instance of \emph{physical access} to a workstation. Subsequently, the breach expanded laterally to compromise additional accounts, ultimately infiltrating the build system.

%\emph{Code Checkout Compromise}
In another scenario, device compromise occurs through \emph{Code Checkout}. This method primarily aims to entice Software Engineers (SWEs) to execute a project on their device and hence compromise it, ostensibly for testing software functionality. For example, in the instance of the \emph{Fake zero-day PoC}~\cite{fake-zero-day-pull-23}, attackers created a bogus account, posing as a security researcher offering a Proof of Concept (PoC) for a zero-day vulnerability. Impersonating security researchers, attackers create authentic-looking repositories using profile images, Twitter accounts, and GitHub repositories to lure victims.

Finally, an alternative method involves compromising a device through physical access. A notable example is the CCleaner attack~\cite{ccleaner-attack}, where the breach originated from an incident of \emph{physical access} to a workstation. Following this breach, the compromise expanded laterally, affecting additional accounts and ultimately compromising the build system.

\subsubsection{\textbf{Malicious Pull Requests}}
Attackers can manipulate contributors into approving their malicious contributions to open-source projects. Hosted on source control systems like GitHub, these contributions manifest as PRs. In a controlled experiment, researchers demonstrated that complex vulnerabilities can deceive even skilled software engineers. They submitted disguised PRs to the mailing list using the standard Linux contribution process. These concealed flaws introduced dormant vulnerabilities, such as a Use-After-Free (UAF) hidden behind apparent legitimacy, employing obfuscation and unconventional pathways. The proposed malicious changes were approved~\cite{wu2021feasibility}. In the \emph{StatCounter} breach, employing code obfuscation method was used to impede developers from easily detecting malicious code during PR reviews~\cite{statcounter2018}. Numerous instances have been documented where malicious code has infiltrated open-source projects. For instance, a joint study by GitHub and Microsoft Research \cite{gonzalez2021anomalicious} pinpointed 15 repositories containing at least one malicious commit. 
%Adversaries exploited email communication, targeting nascent vulnerabilities lacking all requisite conditions. 

%In the \emph{StatCounter} breach, attackers placed packed malicious code in the middle of the code hoping to make it harder for developers to detect malicious code~\cite{statcounter2018}.

%\emph{Big Busy Pull Request} Small pull requests (PRs)
Another form of this attack involves utilizing large PRs, especially auto-generated sections, to bypass reviewer scrutiny and introduce malicious code. Auto-generated files, like npm's lockfiles for dependency installation, are susceptible to manipulations from executed commands, making them prone to malevolent changes. The intricate and frequent alterations they undergo during each PR pose challenges for a thorough review, as emphasized in \cite{snyk-lockfile-19}.

%Other instances of malware introduced by maintainers include the inclusion of "Encryptor" in AWS Terraform modules, the tainted Tasmota firmware for ESP8266 and ESP32 microcontrollers capable of blocking devices, as well as the addition of harmful code to popular NPM packages like "colors" and "faker," affecting a vast number of projects.

%These cases underscore the critical need for enhanced security measures and scrutiny within open-source communities to guard against potential social engineering attacks initiated by those in positions of authority.
% https://avleonov.com/2022/05/11/malicious-open-source-the-cost-of-using-someone-elses-code/

A distinct form of this attack involves submitting a Pull Request to a target project, introducing a controlled \emph{malicious dependency}. In such cases, the dependency may not be inherently malicious but potentially turn malicious in the future. The Agama Wallet incident illustrates this attack, where a pull request added the \emph{electron-native-notify} package. Although initially benign, the package later turned malicious, enabling attackers to steal wallet seeds and other login passphrases used in the application~\cite{snykblog, agama-electron-hack}.

%Similarly, the JavaScript package "get-cookies" targeted unsuspecting applications through a fraudulent pull request. The package contained a backdoor, permitting remote code execution. The npm security team responded quickly, but trust-building tactics like mimicking credible names, inflated download count, and indirect dependencies deceived users to ~\cite{getcookies, getcookiespr}.

In more sophisticated attacks, attackers might focus on a dependency of a dependency or an indirect dependency. For instance, in the event-stream attack~\cite{eventstream-hack}, attackers aimed at a bitcoin wallet platform by exploiting its reliance on the ``event-stream'' package, which in turn depended on ``flatmap-stream.'' The actor ``right9ctrl'' gained the trust of event-stream maintainers, convincing them to merge a new package - flatmap-stream - into event-stream. Initially innocuous, this contribution later became malicious and triggered an attack. Flatmap-stream assessed explicitly if it was executed during building the Copay app~\cite{eventstream-hack}.

Finally, attackers may leverage social engineering tactics through pull requests (PRs) to communicate with developers, potentially embedding phishing links within the PR description, as demonstrated in the work by Siadati et al.\cite{siadati2017x}.

\subsubsection{\textbf{Malicious Dependency Watering hole}} Within this type of attack, perpetrators craft and disseminate malicious packages, anticipating their use by unsuspecting individuals. In the prevalent form known as Typosquatting, developers inadvertently fall prey to the resemblance in names between a malicious component and a genuine dependency. To achieve this, the attackers deliberately select names for their malicious packages that closely mimic those of widely recognized and trusted counterparts. This strategy, also called Typosquatting,  aims to exploit user typographical errors, causing them to download malicious packages or services mistakenly. Attackers replicate the names of legitimate packages within public repositories, anticipating that users or developers will inadvertently select these fraudulent alternatives. Approximately 61\% of malicious packages utilize Typosquatting by imitating existing package names \cite{ohm2020backstabber}. Several incidents are reported, such as 700 typosquatted libraries On RubyGems\cite{RubyGems-Typosquatted}, PyPI~\cite{Sonatype-Typosquatted, Colourama}, golang~\cite{golang-Typosquatting-poc}, and NPM~\cite{npm-typo-squatting, npm-typo-squatting-HackTask}. 

Typosquatting is a cheap and easily executed tactic that significantly affects numerous developers. In one case, attackers generated eight packages with distinct names but identical functionality~\cite{JFrog-Malicious-PyPI}, and developers unknowingly incorporated them. In this incident, the malicious batch has been downloaded 30,000 times.  Another example is ~\cite{twillio-brandjacking}, which targets Twillio developers and fully compromise their machine. Another attacker distributed the ~\emph{mathjs-min} posing as a minified version of wildely-used JS library~\cite{mathjs-min-hack}. The malicious code was added to an innocuously named commit and deeply embedded in the library's files. Another example is RXDrioder SDK which presents itself as an ad-related SDK~\cite{simbad}. It is believed that 206 application developers were scammed to use this package. As a result of this attack, 150 million devices were infected. Similar attacks are being observed in the Ruby gem ecosystem~\cite{bootstrap-sass-package}. In 2018 alone, research unveiled a cumulative 600 million downloads \cite{package-Typosquatting-21} of typosquatted packages.

Adversaries have easily scaled Typosquatting by automating the cloning of GitHub repositories and infusing them with malevolent code. They adopt names closely mirroring authentic package names, albeit within separate repositories. In a singular instance, the attacker crafted 35,000 packages using this method, intending to compromise a substantial array of software products \cite{github-clone-21}. Some of the GitHub supply chain attack forks had several stars, some dating back around five years, suggesting they had been in active use.

Brandjacking represents another form of this attack, involving the distribution of a malicious package with an identical name to a well-known one. A significant incident occurred in 2019 against iOS, targeting developers~\cite{XCodeGhost}. This resulted in injecting malicious code into 350 apps, including WeChat, impacting hundreds of millions of users. In another instance, attackers created \emph{web-browserify} to target developers using the well-known \emph{browserify} package, used for developing cross-platform Node.js-style modules~\cite{sonatype-brandjack}. 

%These attacks involve legitimate users being deceived by the naming of the trusting packages and adding a malicious dependency to their projects. This could happen due to legitimate-looking dependencies.  In 2018 alone, research unveiled over 100 malicious packages, amassing a cumulative 600 million downloads \cite{package-Typosquatting-21}.

\subsubsection{\textbf{Malicious Code Snippet Watering hole}} This attack occurs when an attacker circulates vulnerable sample code, hoping others will incorporate it, making their applications vulnerable. Research has identified numerous vulnerable code snippets on technical social media platforms like Stack Overflow. Out of 72,483 scrutinized code snippets used in at least one GitHub-hosted project, 99 were found to be vulnerable. These compromised snippets, discovered on Stack Overflow, were used in 2,859 GitHub projects \cite{verdi2020empirical}. Although no study has determined whether these instances result from well-intentioned but misguided suggestions or intentional actions, it is conceivable that attackers might exploit this method to propose malicious code intentionally. These attackers could manipulate ranking mechanisms on these platforms to boost the visibility of their harmful code.

\subsubsection{\textbf{Entering the Rank of Maintainers}} In this attack, attackers engage in social engineering to persuade core developers of a project to grant them the role of project maintainer. Open-source project maintainers have significant privileges and can unilaterally modify the code or project settings. An example of this attack is the compromise of the widely-used event-stream package (with 1.9 million weekly downloads on NPM). In this incident, the attacker, having gained the trust of the repository owner, obtained privileged access to the repository and injected a backdoor into the code \cite{eventstream-hack, event-stream-vulnerability}. The social engineering aspect involved building trust with the repository owner by offering assistance to add a specific feature.

Certain maintainers, motivated by financial or political interests, act maliciously in a different form of this attack. An instance includes a maintainer turning rogue for political reasons, releasing a malicious version of the "node-ipc" library with obfuscated code~\cite{node-ipc-malicious-maintainer-21}. In some border cases, like the SSH Decorator (ssh-decorate) incident, it remains unclear whether the intentional code vulnerability resulted from a developer going rogue or if their account was compromised.

\section{Analysis}
\label{ch-analysis}

As outlined in section~\ref{ch-soce-techniques}, real-world SSC attacks encompass six primary social engineering types. We counted attack types in the collected dataset of SSC attack incidents. We observe that approximately 27\% (53 out of 198) of SSC attack reports mentioned at least one social engineering type used by attackers. The distribution of attack types is presented in Table~\ref{tab:attacks-disct}. The most prevalent type of attack is the \emph{Malicious Dependency Watering Hole}, with Typosquatting being the primary subcategory. The second most common type is \emph{Malicious Pull Request}.

\begin{table*}[htbp]
  \centering
  \caption{Occurrence of each type of DevPhish in Software Supply Chains Incidents}
  \begin{tabular}{|l|c|c|}
    %\toprule
    \hline
    \textbf{Attack Type} & \textbf{Number of Incidents} & \textbf{\%} \\
    %\midrule
    \hline
    Account Compromise & 4 & 7.5\% \\
    Device Compromise & 5 & 9\% \\
    Malicious Pull Requests & 7 & 13\% \\
    Malicious Dependency Watering hole & 33 & 62\% \\
    Malicious Code Snippet Watering hole & 1 & 2\%\\
    Entering the Rank of Maintainers & 3 & 6.5\% \\
    %\bottomrule
    \hline
    Total & 53 & 100\%\\
    \hline
  \end{tabular}
  \label{tab:attacks-disct}
\end{table*}

\section{Discussions} 

Although the cases examined in this paper characterize DevPhish as a current threat to SSC security, it is crucial to explain the lessons learned from these threats, their impact, and how the security community should respond to emerging threats in the future. This section delves into the implications of our investigation and provides recommendations for potential paths forward.

%While the cases discussed in this paper qualify DevPhish as a contemporary threat to SSC security, what can be learned from these threats and their impact and how the security community should react in the future to new emerging threats need to be clarified. In this section, we discuss the implications of our investigation and recommend potential routes forward.

\noindent \textbf{Community Consensus.} 
Establishing robust auditing mechanisms on a large scale to shield SSC from DevPhish attacks is a complex undertaking that demands community consensus on the precise places and methods for implementing these auditing measures. The absence of auditing mechanisms in the intricate landscape of the modern SSC ecosystem creates opportunities for DevPhish attacks.

An emerging approach involves auditing build and publish attestations, exemplified by initiatives like those offered by NPM and the Sigstore project~\cite{npm-attestation-2023}. This technique aims to identify discrepancies during a phishing attack. For instance, if an attacker compromises a software repository account and uploads packages directly without committing the corresponding code changes to GitHub repositories, it becomes more challenging for the attacker to erase all traces.    

%\ak{hossein, can describe one or two lines on the commit/push validation and topics to reduce the attack surface. }

However, enforcing the mechanism at scale is a multi-dimensional problem. We understand that finding the right incentives for developers, code maintainers, and repositories to
implement verification policies might be challenging on a global scale,
as these entities might be wary of policing the development community.
However, a coalition in this direction could be helpful as today Sigstore~\cite{sigstore-2023} is the main or perhaps the only line of
defense for protecting software systems from these attacks.
%\ak{please fill the placeholder above.}

\noindent \textbf{Updating Threat Models.} Our analysis pipeline incorporates multiple vantage points to look
at the DevPhish threats. It is critical to ensure the SSC community considers all the possible attack surfaces, including human factors. 
There have been several research on different aspects of the SSC ecosystem and  different security systems that have been developed to characterize the threat and issues (e.g., software fuzzing methods, taint tracking). However, attackers' capabilities to influence the development lifecycle are not studied well. Given the number of incidents as well as the consequence of DevPhish attempts, it is critical to make developers more aware of adversaries' capabilities with a more comprehensive list of possible threats, including human factors.

\noindent \textbf{Robustness and Generalizability.} In our evaluation, we rely on the public account of an incident to determine the presence or absence of social engineering in an attack. It is possible that certain social engineering aspects may not be fully reflected in the report, potentially affecting our assessment of social engineering involvement in SSC attacks. Therefore, we regard our measurement as a conservative estimate.

Incomplete reports posed challenges in categorizing certain attack cases due to limitations in the available data. For instance, it remains unclear whether the SSH Decorator (ssh-decorate) incident resulted from the developer going rogue or their account being compromised. Additionally, some cases involved multiple social engineering attacks, such as device compromise followed by account compromise. In our assessment, we only consider the occurrence at the initial point of impact. 

The attack reports we utilized grouped instances of attacks, potentially based on attack campaigns. For instance, Typosquatting cases were mainly categorized according to the target packages they impersonate, without accounting for the number of successful instances of SSC compromise resulting from them.

%grouping the attacks ... 

%TODO: talk about SDLC steps with more compromise compared to others 

\section{Related Work}
This work lies at the intersection of \emph{software supply chain security} and \emph{social engineering}. For background on software supply chain security, we refer the reader to Section~\ref{ch-ssc-security}. Numerous studies have explored how to secure the software supply chain~\cite{ssdf-2024, esf-2024}. These methods aim to prevent or detect attacks at various stages~\cite{slsa} (see Figure~\ref{fig:slsa}) of the software supply chain, including \emph{Source}, \emph{Build}, \emph{Dependencies} usage, and \emph{Package} distribution. A key concept in preventing software supply chain attacks is software artifact \emph{Provenance} checks. This involves verifying the origin of software at different levels to ensure they are produced by authorized systems and individuals. In-toto~\cite{intoto-2024} is a framework that defines the expected sequences of software provenance in a supply chain and enables attestations that verify required software integrity constraints. However, these checks are ineffective if a developer is socially engineered to approve an attacker's actions. Similarly, several comprehensive surveys have examined attack techniques targeting software supply chains~\cite{ladisa2023sok, ohm2020backstabber, sscsconatiners2023}, yet none have specifically focused on the social engineering aspect of these attacks. This work addresses this gap by focusing on how social engineering can bypass existing checks and calls for further research on this prevalent issue.

\section{Conclusion}
Social engineering is a factor in approximately 27\% of analyzed SSC attacks, encompassing six identified categories, many of which differ from traditional social engineering attacks. This highlights the need to develop custom prevention and detection mechanisms tailored to the workflows of software developers. We are sharing our compiled dataset of SSC attack incidents with the community, encouraging additional exploration and research in this domain.

%\section{Acknowledgment}
%We express our gratitude to Dr. Trishank Karthik Kuppusamy and Dr. Amin Kharraz for their invaluable feedback and contributions to this draft.



%\bibliographystyle{IEEEtranS}
%\section{References}
\bibliographystyle{IEEEtranS}
\bibliography{ref-file}

\end{document}